\documentclass{PoS}

\title{LOFAR-UK}

\ShortTitle{LOFAR-UK}

\author{\speaker{Rob Fender}\\
        School of Physics \& Astronomy, University of Southampton\\
        E-mail: \email{r.fender@soton.ac.uk}}

\author{The Universities of Aberystwyth, Birmingham, Cambridge,
  Cardiff, Durham, Edinburgh, Glasgow, Hertfordshire, Leicester,
  Liverpool John Moores, Kent, Manchester, Newcastle, Nottingham, Open
  University, Oxford, Portsmouth, Queen Mary, University of London
  (QMUL), Sheffield, Southampton, Sussex, and University College
  London as well as STFC/RAL and the UK ATC}

\author{The South East England Physics Network (SEPnet)}

\author{The Science and Technology Facilities Council (STFC)}

\abstract{The LOFAR-UK station at Chilbolton has recently been
  completed and significantly increases the angular resolution of the
  International LOFAR Telescope, as well as providing a unique
  training site and testbed for british experience with
  next-generation software telescopes. The station has been funded
  primarily through the LOFAR-UK, the largest astronomy collaboration
  in Britain, as well as via the South East Physics Network (SEPNET)
  and STFC. In this brief paper we discuss the history and
  organisation of LOFAR-UK, provide a technical description of the
  Chilbolton site, and discuss how LOFAR stations can be augmented by
  the addition of extra local processing capabilities such as ARTEMIS.}

\FullConference{10th European VLBI Network Symposium and EVN Users Meeting: VLBI and the new generation of radio arrays\\
		September 20-24, 2010\\
		Manchester Uk}

\begin{document}

\section{LOFAR}

LOFAR (www.lofar.org) is a `next generation' radio telescope operating
at the low frequency (30--240 MHz) end of the radio spectrum. The
telescope will be more than two orders of magnitude more sensitive
than previous observatories in this frequency range, a dramatic
increase facilitated in large part by advances in computing power and
data transport, hence the use of the term `software telescope'. LOFAR
is also the largest low-frequency pathfinder for the Square Kilometre
Array (SKA, www.skatelescope.org). The project is led by ASTRON and
most of the collecting area ($\sim 70$\%) is in The Netherlands.

International stations are important for LOFAR both scientifically
because they improve the angular resolution of the telescope, and
politically because they draw international partners into the
project. LOFAR stations separated by large distances also act as
useful anti-coincidence tests against local RFI when making wide-field
searches for fast transients.  On 12 June 2010, at the official
inauguration of LOFAR, the Memorandum of Understanding (MoU) of the
International LOFAR Telescope (ILT) was signed by partners from The
Netherlands, Germany, Sweden, France and The United Kingdom. On 9
November 2010 the ILT Foundation came into existence and LOFAR can
genuinely be said to be a pan-European project, with the UK as one of
its founding members.

\section{LOFAR-UK}

\begin{figure}[h]
\includegraphics[width=.95\textwidth]{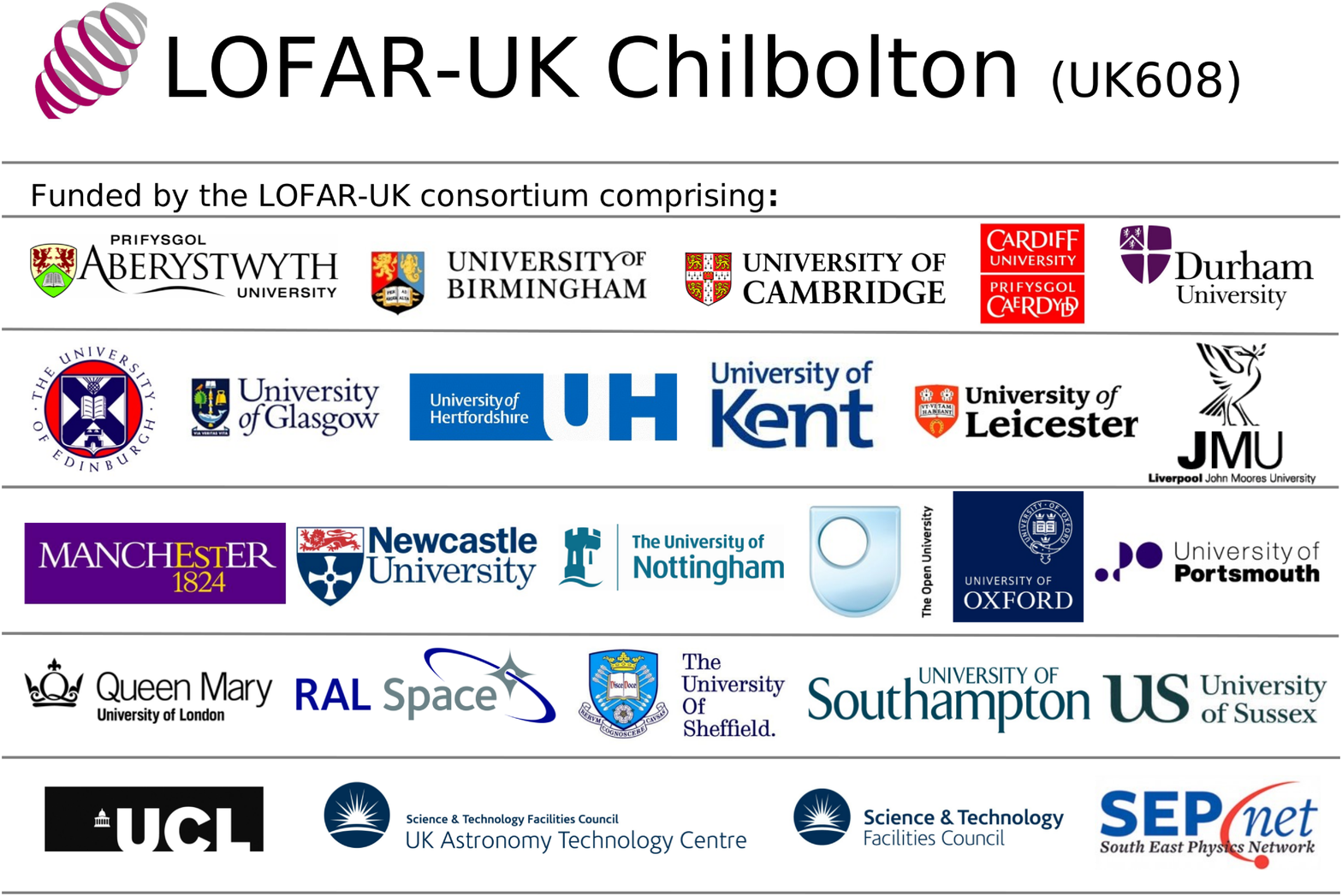}
\caption{The institutional logos of the LOFAR-UK consortium, the
  largest Astronomy consortium in the UK. Credit Martin Bell.}
\label{sign}
\end{figure}

\subsection{Background}

The LOFAR-UK (www.lofar-uk.org) consortium was created in 2006 with
the goal of securing UK involvement in LOFAR. The aim was to purchase,
construct and connect one or more LOFAR stations and in this way allow
UK involvement in LOFAR science, mainly via the existing Key Science
Projects (KSPs). The case for UK participation in LOFAR was presented
in the LOFAR-UK white paper (Best et al., 2008). By 2009 over 20 UK
institutions had joined the project, directly committing resources to
the project (in most cases funding at the level of GBP 50k). The
research council site at Chilbolton (www.chilbolton.rl.ac.uk), in the
south of England, was identified as the prefered site for the station,
although equally good sites at Jodrell Bank and Lord's Bridge were
also considered. Significant funding (GBP $\sim 300$k) was also
provided to the project by the UK South East Physics Network (SEPnet,
www.sepnet.ac.uk), which identified LOFAR as its key astrophysics
theme. Figure \ref{sign} illustrates the breadth of the consortium.

In early 2010, the Science and Technology Facilities Council
(www.scitech.ac.uk) also made a significant contribution (GBP
$\sim200$k) and agreed to cover operations costs in 2011 and 2012. At
this point the contract was signed with ASTRON and construction began
at Chilbolton. At the time of writing (March 2011), construction of
the station is complete, the fibre connection to the correlator in
Groningen is complete, and scientific commissioning of both
international array and `stand-alone' operations has begun.

\subsection{Chilbolton}

The site chosen for the LOFAR-UK station was the Chilbolton
Observatory. This is an existing Science and Technology Facilities
Council (STFC) facility run by the Chilbolton Group of the Rutherford
Appleton Laboratory. The site is located on the edge of the village of
Chilbolton near Stockbridge in Hampshire, England. Apart from the
necessary land area, the site has a good aspect and all the necessary
scientific infrastructure to accomplish the project.

The LOFAR facility installed is what is referred to as an
International Station. It comprises two fields of antennas. They are
the High-Band Array (HBA) and Low-Band Array (LBA), with the high and
low referring to the frequency ranges. The LBA comprises 96 aerials
set out in a pseudo-random pattern with a diameter of approximately
70m. Each aerial stands 1.8m tall, with two crossed, sloped
dipoles. The nominal frequency range of the LBA system is 30 to 80
MHz, although observations have since been successfully carried out
beyond this range, as low as 22 MHz.

\begin{figure}
\includegraphics[width=.95\textwidth]{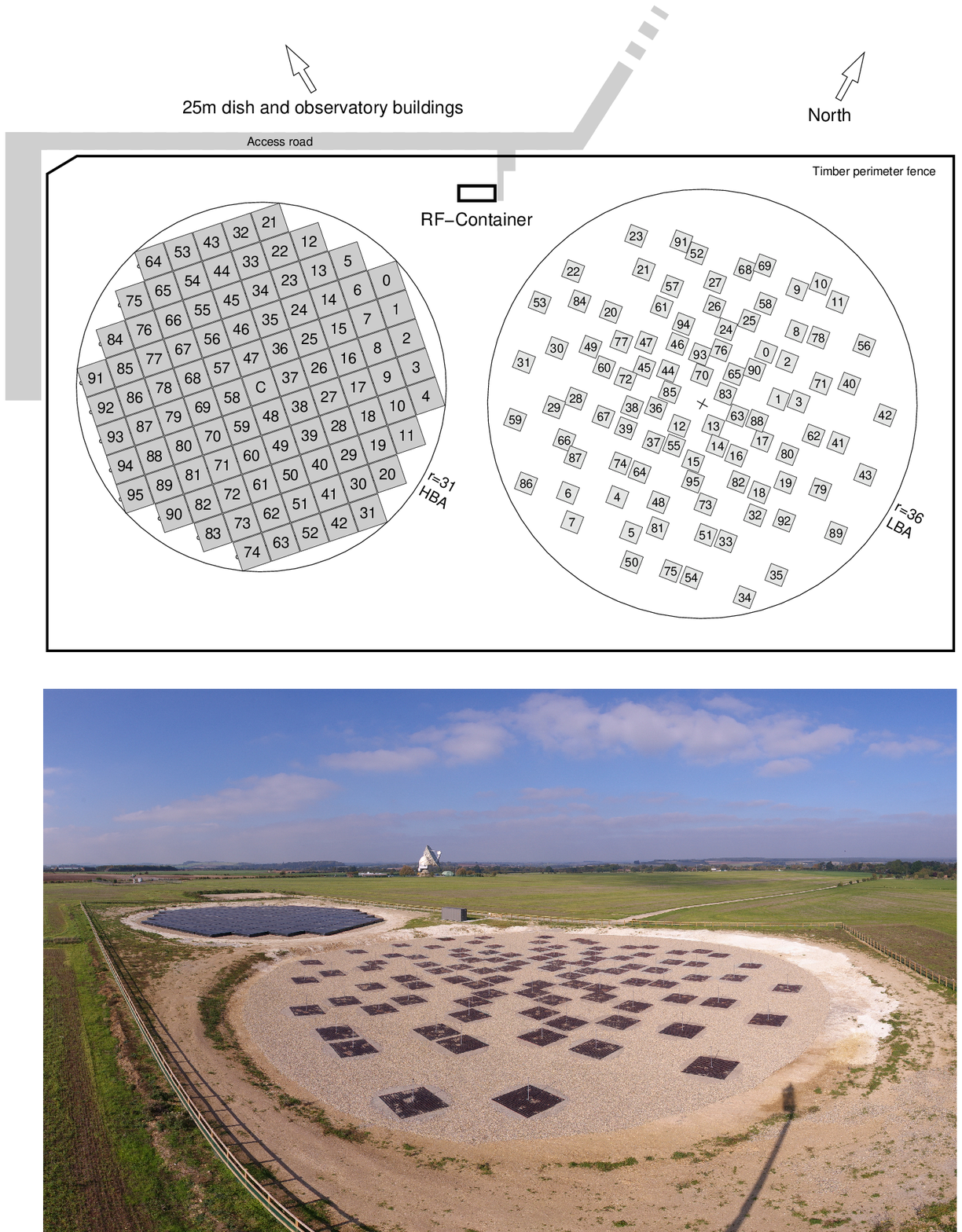}
\caption{Station layout for the LOFAR station at Chilbolton, together
  with a photograph of the completed station. The LBA field is in the
  foreground, with HBA field behind it, and in the distance the
  Chilbolton radio dish, not part of the LOFAR station. The site
  building, including the LOFAR-UK control room (and ARTEMIS), is
  close to the dish. Credit STFC.}
\label{chil}
\end{figure}

The HBA comprises 96 square tiles, 5 by 5 metres in size with a height
of 0.5m. Each tile is split into 16 cells containing crossed-bowtie
antennas. Each tile has its own analogue beamformer to combine the
signals from the cell antennas. The outputs of the 96 tiles can then
be treated as individual antennas in the same way as the LBA. The HBA
system covers the 120 to 240 MHz range. A 97th passive tile is located
in the centre for RF-balancing and wind-loading reasons; it has no
signal output (it is indicated by a C in Fig \ref{chil}).

The 80 to 120 MHz spectral region is the allocated band for FM radio
broadcast and the LOFAR system is designed to suppress this spectral
region. The antenna fields are enclosed within a perimeter fence, made
of timber and synthetic mesh to minimise the RF impact. The fence
prevents incursions by wildlife or stock onto the antenna field.

Also within the enclosure there is a converted 20-foot (6.1 m)
shipping container with air-conditioned Faraday enclosure holding the
digital signal processing electronics. Two coaxial cables connect each
aerial and tile to the central RF-container. These carry the analogue
RF signals, as well as DC power and low-baud control signals.

From the RF container, there is an optical fibre which carries the
digital data to the main observatory buildings. A further network
interface there completes the connection of the local system to a
commercial fibre routed offsite. This link extends to Groningen in the
Netherlands, which is where the Central Processing (CEP) supercomputer
facility is located.

\subsubsection{Construction history}

Installation work was carried out during the summer of 2010. The
project management, site supervision and quality assurance was carried
out by STFC staff, with the primary contractors for the ground works
being Kier-Moss and subcontractors Coral Constructors. Antenna
hardware was supplied by Autonational and RF electronics installation
was carried out by Excel assemblies (both under contract from
ASTRON). After an initial planning phase, ground works commenced on 17
March 2010. In early June, the LBA aerials were installed with the
support of students of local universities. The following month, the
HBA tiles were installed with assistance from U. Manchester and
U.Cambridge. A SEPnet data acquisition system, developed and installed
by U.Oxford (see section 2.3) demonstrated a working system in early
September and the facility was officially opened by Dame Jocelyn Bell
Burnell on 20 September 2010. The station's official designation is
UK608.

Construction and project management at Chilbolton was led by Derek
McKay-Bukowski (SEPnet / STFC).

\subsubsection{Timeline}

The following is a timeline of important events in the construction
and commissioning of UK608 Chilbolton.

\begin{table}[h]
\begin{tabular}{ll}
17-Mar-2010 & Start of ground works\\
19-Apr-2010 & RF-container placed in final location\\
18-May-2010 & Laying of all antenna signal cables complete\\
07-Jun-2010 & Start of LBA aerial deployment\\
11-Jun-2010 & LBA installation complete\\
02-Jul-2010 & Start of HBA tile deployment\\
14-Jul-2010 & HBA installation complete\\
15-Jul-2010 & Electronics in RF container installed\\
01-Sep-2010 & First signal from antenna to fibre\\
06-Sep-2010 & First pulsar detected using local computing\\
20-Sep-2010 & Site officially opened by Dame Jocelyn Bell Burnell\\
12-Oct-2010 & Network installation and offsite fibre\\
09-Dec-2010 & First network packet at 10 Gbit/s from site to CEP\\
11-Jan-2011 & First fringes between UK608 and Dutch core stations\\
18-Jan-2011 & Observations for first long-baseline image\\
\end{tabular}
\end{table}

\begin{table}[h]
\caption{Low band array -- technical details}
\begin{tabular}{ll}
Antenna:	& 96 sloped crossed-dipoles\\
Polarisations:	& 2 linear\\
Longitude:	& 1.433500703 deg W\\
Latitude:	& 51.143833512 deg N\\
Mean height:	& 176.028 m\\
Diameter:	& 70 m (approx)\\
\end{tabular}
\end{table}

\begin{table}[h]
\caption{High band array -- technical details}
\begin{tabular}{ll}
Array:		& 96 tiles\\
Tile:		& 16 crossed-bowtie\\
Polarisations:	& 2 linear\\
Longitude:	& 1.434454726 deg W\\
Latitude:	& 51.143546200 deg N\\
Mean height:	& 177.048 m\\
Diameter:	& 62 m (approx)\\
\end{tabular}
\end{table}

\subsection{Artemis}

\begin{figure}
\includegraphics[width=.95\textwidth]{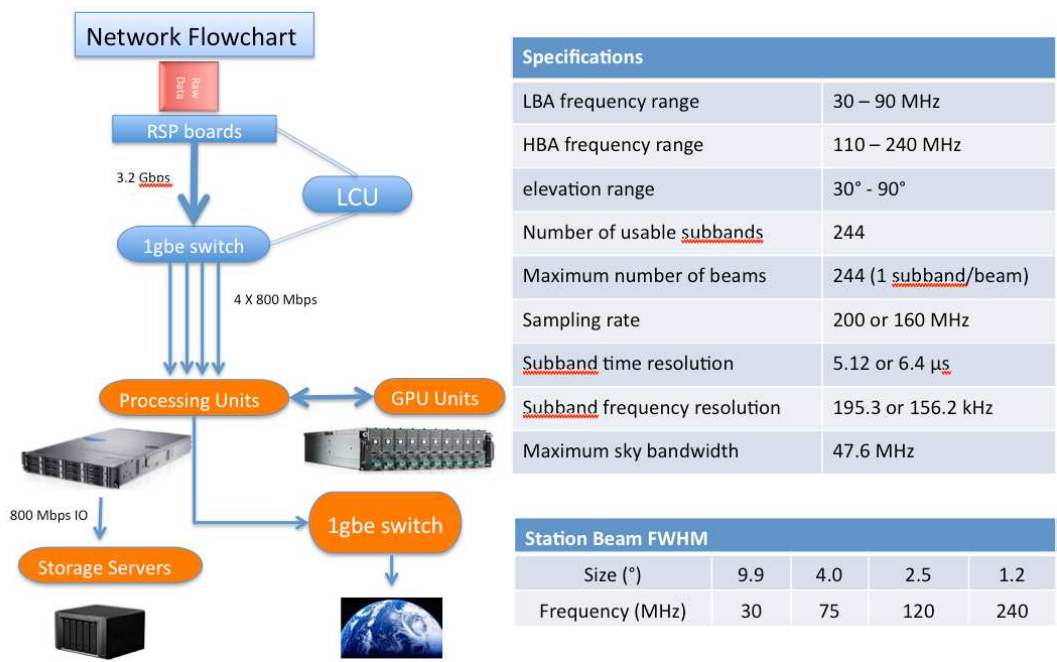}
\caption{A flow diagram for ARTEMIS. Credit Aris Karastergiou.}
\label{artemis}
\end{figure}

During the last year, a team from Oxford Astrophysics and the Oxford
e-Research Centre, has put together an instrument capable of
processing the beamformed data from international LOFAR stations, with
the main objective to conduct large field of view surveys for fast
radio transients. The instrument, code-named ARTEMIS (Advanced Radio
Transient Event Monitor and Identification System, and ancient Greek
goddess of hunting), has been demonstrated to successfully process the
full 3.2 Gbps streaming bandwidth generated by the LOFAR station in
Chilbolton UK.  Fig \ref{artemis} presents a flow diagram for ARTEMIS.
This flow of data is split into four 800 Mbps streams, each of which
is fed into a high performance server, connected to eight high-end
NVIDIA {\em Tesla} graphics cards (C1060). The servers are capable of
reading, and processing the data in real-time, with the possibility to
flag interesting events and also write out to disk or to a network
stream. The processing involves RFI excision, detection, integration
and dedispersion, which are all separate parts of a modular software
pipeline.

The ARTEMIS backend can behave both as a general purpose backend for
the science targets mentioned above, and specifically for the
detection of transients - the latter can also operate commensally with
other possible LOFAR observations. Furthermore, ARTEMIS is telescope
agnostic; it was designed and built with LOFAR in mind, but can be
easily adapted to fit other instruments. For example the newly
constructed EMBRACE aperture arrays, which even share some of the
electronic component design of LOFAR can be exploited by fitting an
ARTEMIS backend to record or process data on the fly. Both EMBRACE and
the LOFAR arrays are pathfinders for the SKA, and ARTEMIS is designed
as an expandable solution.

The ARTEMIS project is led by Aris Karastergiou (Oxford).

\section{The future}

At the time of writing Chilbolton has completed its first test
observations as both a stand-alone station and as part of the ILT (see
timeline above). The coming year will also be mainly devoted to
testing and commissioning, with a number of further stand-alone
observations (including pulsars, SETI work, solar observations,
searches for bright, rare transients) and long-baseline tests as part
of the ILT. In the very near future STFC will sign bilateral agreement
for operations with ASTRON, on behalf of LOFAR-UK. Many tens of
british scientists are already involved in analysis of LOFAR data, and
this number is growing fast.

The LOFAR-UK collaboration should, furthermore, act as an indicator of
the strong and diverse interest in `next generation' radio astronomy
across the UK. This can become a model for how we build national
interest in a strong and energetic scientific community for
participation in the Square Kilometre Array.

\section*{Acknowledgements}

LOFAR-UK would like to acknowledge the invaluable work provided,
mostly for free, during the construction of the LOFAR-UK station at
Chilbolton by an army of PhD students, postdocs and staff.

\section*{References}

\noindent
Best P., and the LOFAR-UK consortium, 
{\em LOFAR-UK White Paper: A Science case for UK involvement in LOFAR},
{\bf arXiv:0802.1186}

\end{document}